\documentclass[11pt,a4paper]{article}

\usepackage{epsfig,amsthm,amscd,amsxtra,amsbsy}
\usepackage{amsmath,amssymb,latexsym, amsfonts}

\begin{document}

\newcommand{\bp}{{\mbox{}\hskip 0.1cm}}
\newcommand{\ba}{\begin{equation}}
\newcommand{\ea}{\end{equation}}

\newtheorem{theorem}{Theorem}[section]
\newtheorem{lemma}{Lemma}[section]
\newtheorem{proposition}{Proposition}[section]
\newtheorem{definition}{Definition}[section]
\newtheorem{assumption}{Assumption}[section]

\textwidth 180mm
\textheight 210mm

\baselineskip 0.6cm
\parskip 0.0cm
\parindent 0.7cm

\pagenumbering{arabic}

\title{\bp { A Recurrence Formula  for Solutions of Burgers Equations }\thanks{
\quad Supported by the Special Funds for Major State Basic
Research Projects , G 1999, 032800}}
\author{Hongling Su$^1$\qquad
Mingliang Wang$^{2}$\qquad Mengzhao Qin$^1$ \\
$^1$ CAST (World Laboratory), \\
Institute of Computational Mathematics\\
 and Scientific/Engineering Computing, \\
Academy of Mathematics and System Sciences,\\
Chinese Academy of Sciences, Beijing, 100080, China\\
$^2$LanZhou University}
\maketitle
\begin{abstract}
A B$\ddot{a}$cklund transformation(BT) and a recurrence formula are derived by the
 homogeneous balance(HB) method. A  initial problem of Burgers equations
 is reduced to a initial problem of heat equation by the BT, the initial problem
of heat equation is resolved by the Fourier
 transformation method, substituting various solutions of  the initial problem
 of the heat equation will yield solutions of the initial problem of the Burgers
equations.
\end{abstract}

\section{Introduction}

    The HB method is used to find solitary wave solutions
and other kinds of exact solutions of nonlinear partial differential equations
(PDEs)$^{[1,2]}$, in this paper, the HB method is used  first time with
the help of the Fourier transformation method to solve some initial problems
of Burgers equations which have been widely used in physical system.

    This paper falls into 3 parts. In section 2, by the HB method we obtain
a general BT$^{3,4}$ which shows both the connection between solutions of Burgers
equations itself  and connection between solutions of Burgers and heat
equations, its special case is Cole--Hopf transformation and especially through the BT
 a recurrence
formula is generated for solutions of Burgers equations as
\begin{align*}
&u_{N+1}=\frac{u_{Nx}+v_{Nx}}{u_N+v_N}+u_N, \\
&v_{N+1}=\Large\frac{u_{Ny}+v_{Ny}}{u_N+v_N}+v_N.
\end{align*}
A reasonable
initial problem of the Burgers equations is put out and reduced to a initial
problem of the heat equation which is solved by Fourier transformation method
in Sec.3. The solution of the initial problem of the heat equation with the aid
of the Burgers yields the solution of the initial problem of the Burgers.

\section{Burgers Equations}

\begin{align}
\begin{split}
&u_t=u_{xx}+u_{yy}+2uu_x+2vu_y, \\
&v_t=v_{xx}+v_{yy}+2uv_x+2vv_y.
\end{split}
\end{align}
  This is the famous Burgers equations which have got much attention in recent
years. In Ref.[7,8,9], high dimensional  and high degree Burgers equations
are studied, and a pair of exact solution of Burgers equations(1) is obtained in
Ref.[4]. In this section we will present some BTs of Burgers equations(1)
through the HB method.

    According to the idea of HB method, we seek for its solution of the form:
\begin{align}
\begin{split}
&u=\frac{\partial f(\varphi)}{\partial x}+u_0=f^{\prime }(\varphi)
\varphi_x+u_0, \\
&v=\frac{\partial g(\varphi)}{\partial y}+v_0=g^{\prime }(\varphi)
\varphi_y+v_0,
\end{split}
\end{align}
here
$f=f(\varphi)$ and $g=g(\varphi)$ which are functions of single variable, and
$\varphi=\varphi(x,y,t)$ are all to be determined later.
$(u_0,v_0)$ are a pair of solutions of $(1)$.
By (2), it is easy to deduce that
\begin{align}
\begin{split}
&u_t=f^{\prime \prime }\varphi_x\varphi_t+f^{\prime }\varphi_{xt}+u_{0t}, \\
&u_{xx}=f^{\prime \prime \prime }\varphi_x^3+3f^{\prime \prime }\varphi_x\varphi_{xx}+f^{\prime
}\varphi_{xxx}+u_{0xx}, \\
&u_{yy}=f^{\prime \prime \prime }\varphi_x\varphi_y^2+2f^{\prime \prime
}\varphi_{xy}\varphi_y+f^{\prime \prime }\varphi_x\varphi_{yy}+f^{\prime}
\varphi_{xyy}+u_{0yy}, \\
&2uu_x=2f^{\prime }f^{\prime \prime }\varphi_x^3+2f^{\prime 2}\varphi_x\varphi_{xx}+2f^{\prime
\prime }\varphi_x^2u_0+2f^{\prime }\varphi_{xx}u_0+2f^{\prime
}\varphi_xu_{0x}+2u_0u_{0x}, \\
&2vu_y=2g^{\prime }f^{\prime \prime }\varphi_x\varphi_y^2+2g^{\prime }f^{\prime
}\varphi_y\varphi_{xy}+2f^{\prime \prime }\varphi_x\varphi_yv_0+2f^{\prime }\varphi_{xy}v_0+2g^{\prime
}\varphi_yu_{0y}+2vu_{0y},
\end{split}
\end{align}
\begin{align}
\begin{split}
&v_t=g^{\prime \prime }\varphi_y\varphi_t+g^{\prime}\varphi_{yt}+v_{0t}, \\
&v_{xx}=g^{\prime \prime \prime }\varphi_y\varphi_x^2+2g^{\prime \prime
}\varphi_x\varphi_{xy}+g^{\prime \prime }\varphi_y\varphi_{xx}+g^{\prime
}\varphi_{yxx}+v_{0xx}, \\
&v_{yy}=g^{\prime \prime \prime }\varphi_y^3+3g^{\prime \prime }\varphi_y\varphi_{yy}+g^{\prime
}\varphi_{yyy}+v_{0yy}, \\
&2uv_y=2f^{\prime }g^{\prime \prime }\varphi_x^2\varphi_y+2f^{\prime }g^{\prime
}\varphi_x\varphi_{xy}+2g^{\prime \prime }\varphi_x\varphi_yu_0 +2g^{\prime }\varphi_{xy}u_0+2f^{\prime
}\varphi_xv_{0x}+2u_0v_{0x}, \\
&2vv_y=2g^{\prime }g^{\prime \prime }\varphi_y^3+2g^{\prime 2}\varphi_y\varphi_{yy}+2g^{\prime
\prime }\varphi_y^2v_0+2g^{\prime }\varphi_{yy}v_0+2g^{\prime
}\varphi_yv_{0y}+2v_0v_{0y},
\end{split}
\end{align}
substituting (3) and (4) into (1), first setting the coefficients of
$\varphi_x^3$ and $\varphi_y^3$ to zero, we obtain ODEs
\begin{align}
\begin{split}
&f^{\prime \prime \prime }+2f^{\prime }f^{\prime \prime }=0, \\
&g^{\prime \prime \prime }+2g^{\prime }g^{\prime \prime }=0,
\end{split}
\end{align}
which have solutions
$f=g=\ln \varphi$, thereby it holds that \\
\begin{align}
\begin{split}
&f^{\prime \prime \prime }+2g^{\prime }f^{\prime \prime }=0, \quad g^{\prime
\prime \prime }+2f^{\prime }g^{\prime \prime }=0, \\
&f^{\prime \prime }+f^{\prime 2}=0, \quad g^{\prime \prime }+g^{\prime 2}=0,
\end{split}
\end{align}
by using (3)--(6), we obtain the expressions:
\begin{align}
\begin{split}
&u_t-u_{xx}-u_{yy}-2uu_x-2vu_y \\
&=f^{\prime \prime }(\varphi_x\varphi_t-\varphi_x\varphi_{xx}-
\varphi_x\varphi_{yy}-2\varphi_x\varphi_yv_0-2\varphi_x^2u_0) \\
&+f^{\prime}(\varphi_{xt}-\varphi_{xxx}-\varphi_{xyy}-2\varphi_{xx}u_0-2\varphi_x
u_{0x}-2\varphi_{xy}v_0-2\varphi_yu_{0y}),
\end{split}
\end{align}
\begin{align}
\begin{split}
&v_t-v_{xx}-v_{yy}-2uv_x-2vv_y \\
&=f^{\prime \prime }(\varphi_y\varphi_t-\varphi_y\varphi_{yy}-\varphi_y
\varphi_{xx}-2\varphi_x\varphi_yu_0-2\varphi_y^2v_0) \\
&+f^{\prime }(\varphi_{yt}-\varphi_{yxx}-\varphi_{yyy}-2\varphi_{xy}u_0-2\varphi_xv_{0x}
-2\varphi_yv_{0y}-2\varphi_{yy}v_0),
\end{split}
\end{align}
It is easy to see that (2) are solutions of (1), provide that the right hand
sides of (7) and (8) to be zero, hence set
\begin{align}
&\varphi_t-\varphi_{xx}-\varphi_{yy}-2\varphi_xu_0-2\varphi_yv_0=0,\\
&\varphi_{xt}-\varphi_{xxx}-\varphi_{xyy}-2\varphi_{xx}u_0-2\varphi_xu_{0x}-2\varphi_{xy}v_0-2\varphi_yu_{0y}=0,
\\
&\varphi_{yt}-\varphi_{yxx}-\varphi_{yyy}-2\varphi_{yy}v_0-2\varphi_yv_{0y}-2\varphi_{xy}u_0-2\varphi_xv_{0x}=0,
\end{align}
then the right hand sides of (7) and (8) respectively vanish. In fact (10) and
(11) can be reduced into
\begin{align}
&(\varphi_t-\varphi_{xx}-\varphi_{yy}-2\varphi_xu_0-2\varphi_yv_0)_x+2\varphi_yv_{0x}-2\varphi_yu_{0y}=0,
\\
&(\varphi_t-\varphi_{xx}-\varphi_{yy}-2\varphi_xu_0-2\varphi_yv_0)_y+2\varphi_xu_{0y}-2\varphi_xv_{0x}=0,
\end{align}
thereby, (9)--(11) equal that
\begin{align}
\begin{split}
&\varphi_t-\varphi_{xx}-\varphi_{yy}-2\varphi_xu_0-2\varphi_yv_0=0, \\
&u_{0y}-v_{0x}=0,
\end{split}
\end{align}
here
$\varphi_x^2+\varphi_y^2\neq 0$. Substituting
$f=g=\ln \varphi$ into(2), we obtain BT of
Burgers equations (1)
\begin{align}
&u=\frac{\varphi_x}\varphi+ u_0,\\
&v=\frac{\varphi_y}\varphi +v_0,\\
&\varphi _t-\varphi _{xx}-\varphi _{yy}-2\varphi _xu_0-2\varphi _yv_0,\\
&u_{0y}-v_{0x}=0,
\end{align}
 namely,
$(u_0,v_0)$ is a pair of solutions of Burgers equations(1),
and $u_{0y}=v_{0x}$ therefore (2) is  another  pair of solutions of (1)
only if $\varphi=\varphi(x,y,t)$ satisfies linear equation(9). Now set
 $\varphi=u_0+v_0$, (9) is automatically satisfied, additionally,
if $u_{0y}=v_{0x}$ exists in the first pair of solutions $(u_0,v_0)$,
the new solutions $(u,v)$ according to the BT have to satisfy $u_y=v_x$,
the circular use of this $BT$ yields the recurrence formula of Burgers
equations (1) as
\begin{align}
\begin{split}
&u_{N+1}=\frac{u_{Nx}+v_{Nx}}{u_N+v_N}+u_N,\\
&v_{N+1}=\frac{u_{Ny}+v_{Ny}}{u_N+v_N}+v_N.
\end{split}
\end{align}

Here $N=0, 1, 2, \cdot \cdot \cdot $, with $v_{0x}=u_{0y}$.

   In addition, if we take initial solution of Burgers equations (1)
as $u_0=v_0=0$ (the trivial solutions of Burgers equations (1)),
BT (15)--(18) respectively reduce to
\begin{align}
\begin{split}
&u=\frac{\varphi_x}{\varphi},\quad\quad v=\frac{\varphi_y}{\varphi}, \\
&\varphi_t-\varphi_{xx}-\varphi_{yy}=0,
\end{split}
\end{align}
which is usually called Cole--Hopf transformation.
 
\section{ Fourier Transformation}

 Provided the initial value problem of Burgers equations in what follows
\begin{align}
&u_t=u_{xx}+u_{yy}+2uu_x+2vu_y,
\\
&v_t=v_{xx}+v_{yy}+2uv_x+2vv_y,
\\
&u\left| _{t=t_0}\right.=s(x,y),\quad v\left| _{t=t_0}\right. =k(x,y)
\end{align}
where $s(x,y), k(x,y)\rightarrow{0}$ when $|x|, |y|\rightarrow {0}$.

   By using the former result, we can take its solution to be as
\begin{align}
u=\frac{\varphi _x}\varphi +u_0, \quad v=\frac{\varphi _y}\varphi +v_0,
\end{align}
here $(u_0,v_0)$ are constant solutions of(21)--(22) and $\varphi $ satisfies
\begin{align}
\varphi _t-\varphi _{xx}-\varphi _{yy}-2u_0\varphi _x-2v_0\varphi _y=0
\end{align}
 then (21)--(23) can be rewritten into the initial value problem of the
following form
\begin{align}
\begin{split}
&\varphi _t-\varphi _{xx}-\varphi _{yy}-2u_0\varphi _x-2v_0\varphi _y=0,\\
&\varphi \left| _{t=t_0}\right. =f(x,y),
\end{split}
\end{align}
which
\begin{align}
&f(x,y)=\text{e}^{\int_{x_0}^xs_1(x,y_0)\text{d}x+\int_{y_0}^yk_1(x,y)\text{d}y}
\end{align}
$(x_0,y_0)$ are  arbitrary constants, $s_1(x,y)=s(x,y)-u_0,k_1(x,y)=k(x,y)-v_0$.
Then if $u_0,v_0$ are arbitrary constant solutions of equations (1),
 by using of Fourier transformation, we obtain  solutions
of (26)--(27).

     Now introduce the Fourier transformation  simply.
If $f(x,y,t)$ is a function,
$\widehat{f}(c_1,c_2,t)$ is used to denote its Fourier transformation as follow:
\begin{equation}
\widehat {f}(c_1,c_2,t)=F[f(x,y,t)]=\frac 1{2\pi}\int_{-\infty }^\infty
\int_{-\infty }^\infty f(x,y,t)\text{e}^{-\text{i}c_1x-\text{i}c_2y}\text{d}x
\text{d}y
\end{equation} and
\begin{equation}
f(x,y,t)=F^{-1}[\widehat{f}(c_1,c_2,t)]=\frac 1{2\pi}\int_{-\infty }^\infty
 \int_{-\infty}^\infty
  \widehat{f}(c_1,c_2,t)\text{e}^{\text{i}c_1x+\text{i}c_2y}\text{d}c_1\text{d}c_2
\end{equation}
we'd like show its several properties which are used  here:\\
I $\quad$ $F[af+by]=aF[f]+bF[g]$\\
II $\quad$ $F[\frac{\partial^{m}\partial^{n} f }{\partial x^{m}\partial y^n}]
   =(\text{i}c_1)^m(\text{i}c_2)^nF[f]$\\
III$\quad$ $F[f(x-a,y-b,t)]=\text{e}^{-\text{i}c_1a-\text{i}c_2b}F[f(x,y,t)]$\\
IV $\quad$ $F[f(ax,by,t)]=\frac 1 {\left |ab\right |}
\widehat{f}(\frac {c_1} {a},\frac {c_2} {b},t)$\\
V $\quad$ if $(f*g)(x,y,t)=\int_{-\infty}^\infty \int_{-\infty}^\infty
f(x-z_1,y-z_2,t)g(z_1,z_2,t)\text{d}z_1\text{d}z_2$, then
\begin{align*}
&F^{-1}[\widehat{f}\widehat{g}]=\frac1 {2\pi} f*g
\end{align*}
here, $a$ and $b$ are arbitrary constants, m and n are arbitrary integers.
In order to deal with the problem (26), the Fourier transformation is done
for them, it is reduced to a order differential problem as following:
\begin{equation}
\begin{split}
&\frac{\text{d}\widehat{\varphi}}{\text{d}t}+c_1^2\widehat{\varphi}+c_2^2\widehat{\varphi}-2iu_0c_1\widehat{\varphi}-2v_0c_2\widehat{\varphi}=0,\\
&\widehat{\varphi}\left|_{t=t_0}\right.=\widehat{f}(x,y),
\end{split}
\end{equation}
which admits solutions as
\begin{align}
&\widehat{\varphi}(c_1,c_2,t)=\widehat{f}(c_1,c_2,t)\text{e}^{-(c_1^2+c_2^2
-2\text{i}u_0c_1-2\text{i}v_0c_2)(t-t_0)}
\end{align}
set $\widehat{g}=\text{e}^{-(c_1^2+c_2^2
-2\text{i}u_0c_1-2\text{i}v_0c_2)(t-t_0)}$ such that
\begin{align}
&g(x,y,t)=F^{-1}[\widehat{g}(c_1,c_2,t)]=\frac 1 {2(t-t_0)}\text{e}^{-\frac
{(x-2u_0)^2+(y-2v_0)^2}{4(t-t_0)}}
\end{align}
so the solution of (26) is gained
$$
\varphi (x,y,t)=F^{-1}[\widehat{f} \widehat{g}]=\frac 1{4\pi (t-t_0)}
\int_{-\infty }^\infty \int_{-\infty }^\infty
\text{e}^{\int_{x_0}^{x-z_1}s_1(\tau_1,y_{0})\text{d}\tau_1+%
\int_{y_0}^{y-z_2}k_1(x-z_1,\tau_2)\text{d}\tau_2}.
$$
\begin{align}
&\text{e}^{-\frac{(z_1-2iu_0)^2+(z_2-2iv_0)^2}{4(t-t_0)}}\text{d}z_1
\text{d}z_2,
\end{align}
hence $u=\frac{\varphi_x}\varphi +u_0\;,\;v=\frac{\varphi_y}\varphi +v_0$
which are solutions of (21)--(23) also can be got, detail discussion
does not present here.

\end{document}